\documentclass[a4paper,twocolumn,showpacs]{revtex4}

\usepackage[english]{babel}
\usepackage{graphicx}
\usepackage{xspace}
\usepackage{longtable}
\usepackage{amsmath,amssymb}

\DeclareMathAlphabet{\mathbfit}{OT1}{cmr}{bx}{it}

\begin{document}

\title{Magnetic moments of W $\mathbfit{5d}$ in Ca$_2$CrWO$_6$ and Sr$_2$CrWO$_6$
double perovskites}

\author{P.~Majewski}
\email{Petra.Majewski@wmi.badw.de}
\affiliation{Walther-Mei{\ss}ner-Institut, Bayerische Akademie der
Wissenschaften, Walther-Mei{\ss}ner Str.~8, 85748 Garching,
Germany}
\author{S.~Gepr\"{a}gs}
\affiliation{Walther-Mei{\ss}ner-Institut, Bayerische Akademie der
Wissenschaften, Walther-Mei{\ss}ner Str.~8, 85748 Garching,
Germany}
\author{A.~Boger}
\affiliation{Walther-Mei{\ss}ner-Institut, Bayerische Akademie der
Wissenschaften, Walther-Mei{\ss}ner Str.~8, 85748 Garching,
Germany}
\author{M.~Opel}
\affiliation{Walther-Mei{\ss}ner-Institut, Bayerische Akademie der
Wissenschaften, Walther-Mei{\ss}ner Str.~8, 85748 Garching,
Germany}
\author{A.~Erb}
\affiliation{Walther-Mei{\ss}ner-Institut, Bayerische Akademie der
Wissenschaften, Walther-Mei{\ss}ner Str.~8, 85748 Garching,
Germany}
\author{R.~Gross}
\affiliation{Walther-Mei{\ss}ner-Institut, Bayerische Akademie der
Wissenschaften, Walther-Mei{\ss}ner Str.~8, 85748 Garching,
Germany}
\author{G.~Vaitheeswaran}
\affiliation{Max-Planck-Institute for Solid State Research,
Heisenbergstr.~1, 70569 Stuttgart, Germany}
\author{V. Kanchana}
\affiliation{Max-Planck-Institute for Solid State Research,
Heisenbergstr.~1, 70569 Stuttgart, Germany}
\author{A. Delin}
\affiliation{Department of Materials Science and Engineering,
Royal Institute of Technology (KTH), 10044 Stockholm, Sweden}
\author{F.~Wilhelm}
\affiliation{European Synchrotron Radiation Facility (ESRF), 6 Rue
Jules Horowitz, BP 220, 38043 Grenoble Cedex 9, France}
\author{A.~Rogalev}
\affiliation{European Synchrotron Radiation Facility (ESRF), 6 Rue
Jules Horowitz, BP 220, 38043 Grenoble Cedex 9, France}
\author{L.~Alff}
\email{alff@oxide.tu-darmstadt.de} \affiliation{Darmstadt
University of Technology, Petersenstr.~23, 64287 Darmstadt,
Germany}

\date{\today}

\begin{abstract}

We have investigated the magnetic moment of the W ion in the
ferrimagnetic double perovskites Sr$_2$CrWO$_6$ and Ca$_2$CrWO$_6$
by X-ray magnetic circular dichroism (XMCD) at the W $L_{2,3}$
edges. In both compounds a finite negative spin and positive
orbital magnetic moment was detected. The experimental results are
in good agreement with band-structure calculations for
(Sr/Ca)$_2$CrWO$_6$ using the full-potential linear muffin-tin
orbital method. It is remarkable, that the magnetic ordering
temperature, $T_\textrm{C}$, is correlated with the magnetic
moment at the 'non-magnetic' W atom.

\end{abstract}
\maketitle

The double perovskites of the composition $A_2 BB^\prime
\textrm{O}_6$ (with $A$ an alkaline earth, $B$ a magnetic
transition metal ion, and $B^\prime$ a non-magnetic ion) are
interesting materials, both due to their rich physics and their
properties promising for applications in spintronics. Recently,
the double perovskites have attracted renewed interest when a
large room-temperature magnetoresistance was observed in
Sr$_2$FeMoO$_6$ with a Curie temperature $T_\textrm{C}=420$\,K
\cite{Kobayashi:98}. Furthermore, band structure calculations
indicated that the ferromagnetic double perovskites not only have
large $T_\textrm{C}$ but also are half-metals. This immediately
suggests their application as a source of spin-polarized charge
carriers in spintronic devices. In addition to Sr$_2$FeMoO$_6$,
ferrimagnetism with a $T_\textrm{C}$ up to 458\,K has been found
in ceramic and thin film samples of the compound Sr$_2$CrWO$_6$
\cite{Philipp:01,Philipp:03}. Furthermore, this compound also has
been predicted half-metallic by band-structure calculations
\cite{Philipp:03,Jeng:03}. The double perovskite with the highest
$T_\textrm{C}$ known so far is Sr$_2$CrReO$_6$ with
$T_\textrm{C}=635$\,K \cite{Kato:02,Asano:04}. However, according
to new results of density functional theory this compound is not
fully half-metallic due to a strong spin-orbit coupling of Re
\cite{Vaitheeswaran:05}.

For clarifying the nature of magnetic exchange in the double
perovskites, the knowledge on the local magnetic moments on the
$B$ and $B^\prime$ site is important. For Sr$_2$FeMoO$_6$, recent
XMCD measurements showed a spin moment of about
$3\,\mu_\textrm{B}$ for Fe$^{3+}$ ion at the $B$ site.
Interestingly, for the non-magnetic Mo$^{5+}$ ion at the
$B^\prime$ site an \emph{antiparallel aligned} spin moment of
about $-0.3\,\mu_\textrm{B}$ and a small orbital contribution was
found \cite{Besse:02}. This observation is in agreement with a
generalized double exchange or kinetic energy driven exchange
model proposed by Sarma {\em et al}.~\cite{Sarma:00} to explain
the strong ferromagnetic exchange in double perovskites despite
the large distance between the magnetic ions (e.~g.~8.82\,{\AA} for
Sr$_2$CrWO$_6$). Subsequently, extensions of this model have been
used to explain ferromagnetism in magnetic semiconductors and
organic ferromagnets \cite{Kanamori:01,Fang:01}. We recently
showed that this model also can be applied to the $A_2$CrWO$_6$
system \cite{Philipp:03}. In the following we shortly describe the
kinetic energy driven exchange model: For the magnetic ion
Cr$^{3+}$ Hund's splitting is much larger than the crystal field
splitting, and the majority spin $t_{\text{2g}}$ band is filled.
In contrast, at the 'non-magnetic' site W$^{5+}$ the crystal field
splitting is large and Hund's splitting small. As the majority
spin bands at the magnetic site are occupied, kinetic energy gain
can only be obtained by hybridization and the hopping of the
minority spin electrons from the 'non-magnetic' site into the
empty minority spin bands of the magnetic ion. By shifting
electrons from the majority spin band of the 'non-magnetic' site
into the minority spin band, the system can gain energy. As a
result, the charge carriers become strongly polarized, in the
extreme case even half-metallic. Further, at the 'non-magnetic'
site a negative spin magnetic moment develops. An evident check of
the validity of this model is the investigation of the magnetic
moment on the non-magnetic $B^\prime$ site. We note that in this
simple model magnetic order is established by a purely electronic
mechanism. In more sophisticated theories anti-site disorder or
breathing distortions are necessary to stabilize ferrimagnetism
\cite{Solovyev:02}. The relevance of such structural degrees of
freedom still has to be investigated experimentally.

Here, we report both on band-structure calculations on the system
(Sr/Ca)$_2$CrWO$_6$ using the full-potential linear muffin-tin
orbital method, and on XMCD measurements detecting the spin and
orbital magnetic moment at the 'non-magnetic' W atom. Using the
magneto-optical sum-rules of XMCD \cite{Thole:92,Carra:93} spin
and orbital magnetic moments can be calculated separately, thereby
allowing for an experimental test of the kinetically driven
exchange model and the band-structure calculations for
(Sr/Ca)$_2$CrWO$_6$.

The half-metallicity of (Sr/Ca)$_2$CrWO$_6$ as derived from the
simple ionic model described above is supported by  band-structure
calculations using an all-electron full-potential linear
muffin-tin orbital method (FP-LMTO) \cite{Wills:00}. One advantage
in this method is, that no shape approximation of the potential,
wave functions, or charge density is made. Spin-orbit coupling is
included in all calculations. For more details see the theoretical
results of this method for the compound Sr$_2$CrReO$_6$
\cite{Vaitheeswaran:05}.

\begin{figure}[tb]
\centering{%
\includegraphics [width=0.8\columnwidth,clip=,angle=-360]{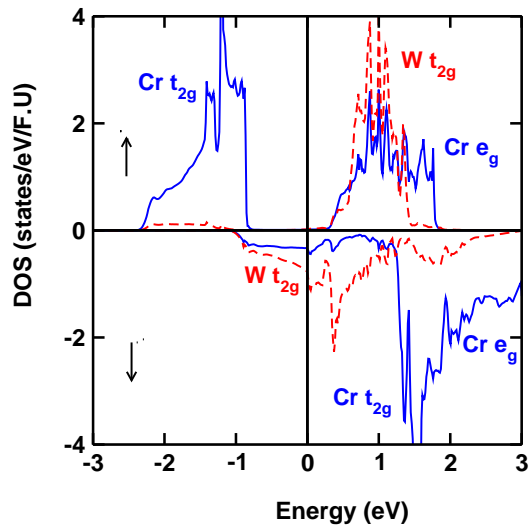}}
 \caption{(Color online) Orbital resolved density of states (DOS)
  of Sr$_2$CrWO$_6$ with spin-orbit coupling
The Fermi energy is indicated by zero energy level.}
 \label{Fig:bandstructure}
\end{figure}

Our results from the band-structure calculations are presented in
Fig.~\ref{Fig:bandstructure}. The spin-up bands, which are plotted
in the upper half of Fig.~\ref{Fig:bandstructure}, show the
crystal field splitting of the Cr $3d$ bands at the Fermi level
whereas the hybridized Cr $3d$ and W $5d$ spin-down bands (plotted
in the lower half) are located at the Fermi level. The magnetic
spin moment at the W site is calculated to be
$m_S=-0.31\,\mu_{\textrm{B}}$ and the ratio of the orbital and
spin moment is found to $|m_{\text{L}}/m_{\text{S}}|=0.32$. The
number of W $5d$ holes deduced from the band-structure
calculations is $n_{\text{h}}=6.3$. Qualitatively, the
band-structure calculated by FP-LMTO appears to be similar to that
calculated by LMTO using the atomic sphere approximation (ASA)
\cite{Philipp:03}. The remarkable point is that half-metallicity
is preserved even if spin-orbit coupling is included. This result
for Sr$_2$CrWO$_6$ is in contrast to that for Sr$_2$CrReO$_6$,
where spin-orbit coupling destroys the half-metallic nature. We
explain this difference between Sr$_2$CrWO$_6$ and Sr$_2$CrReO$_6$
as due to the fact that W has one $5d$ electron less than Re,
causing the W $t_{2g}$ states to shift upwards in energy, away
from the Fermi level. As a result, the hybridization at the gap
becomes less proncounced, and the gap is preserved. In order to
validate the prediction of the band-structure calculations, we
investigated the magnetic moment on the W atom of the double
perovskites Sr$_2$CrWO$_6$ and Ca$_2$CrWO$_6$ by XMCD.

The sample preparation is described elsewhere \cite{Philipp:03}.
In short, the polycrystalline samples contain small amounts of the
parasitic phases W and $A_3$WO$_6$ (with $A=$\,Sr, Ca). It is
important to note that we found no magnetic W impurity phases,
which could possibly influence the XMCD measurements. From SQUID
measurements we obtained $T_\textrm{C}=443$\,K for Sr$_2$CrWO$_6$
and $T_\textrm{C}=160$\,K for Ca$_2$CrWO$_6$.

The XMCD measurements on the W L$_{2,3}$ edges were performed at
the European Synchrotron Radiation Facility (ESRF) at beam line
ID12 \cite{Rogalev:book}. The spectra were recorded using the
total fluorescence yield detection mode. The XMCD spectra were
obtained as direct difference between consecutive XANES scans
(X-ray Absorption Near Edge Spectrum) recorded with opposite
helicities of the incoming X-ray beam. To ensure that the XMCD
spectra are free from any experimental artefacts the data was
collected for both directions of the applied magnetic field of
7\,T (parallel and antiparallel to the X-ray beam). The degree of
circular polarization of the monochromatic X-ray beam was 98\%.
The measurements were performed at low temperature for the
Ca$_2$CrWO$_6$ sample ($T<T_{\text{C}}$) and at room temperature
for the Sr$_2$CrWO$_6$ sample ($T<T_{\text{C}}$).

Since the samples measured in backscattering geometry were very
thick, the spectra were first normalized to the edge jump of unity
and then corrected from self-absorption effects.  The edge jump
intensity ratio $L_3/L_2$ was then normalized to 2.19/1
\cite{Wilhelm:01}. This is different from the statistical 2:1
branching ratio due to the difference in the radial matrix
elements of the 2p1/2 to 5d(L2) and 2p3/2 to 5d(L3) transitions.

\begin{figure}[tb]
\centering{%
\includegraphics[width=0.9\columnwidth,clip=]{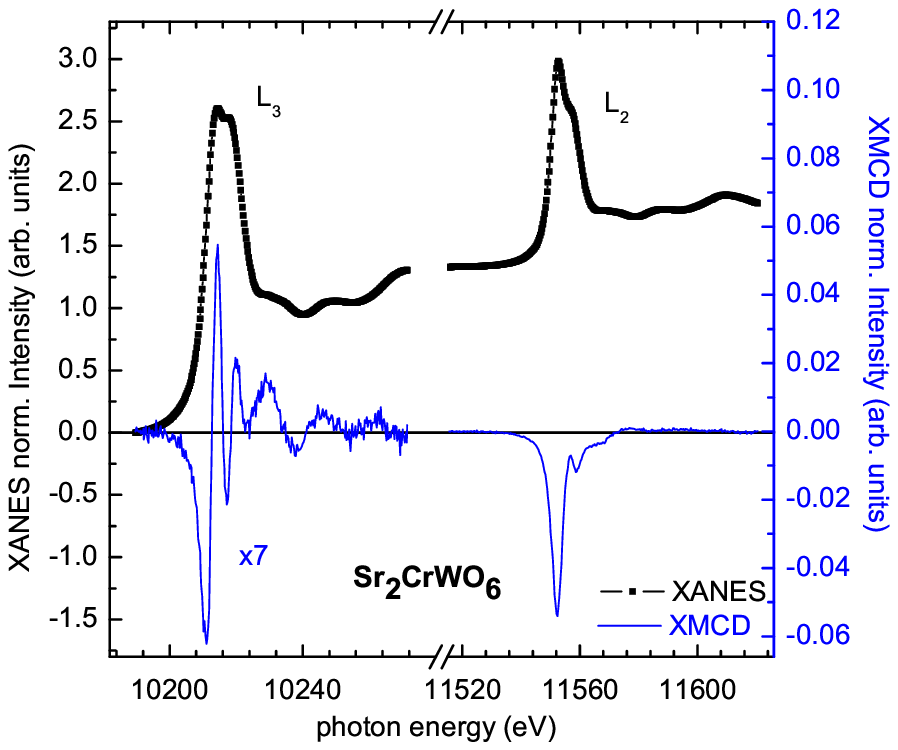}
\includegraphics[width=0.9\columnwidth,clip=]{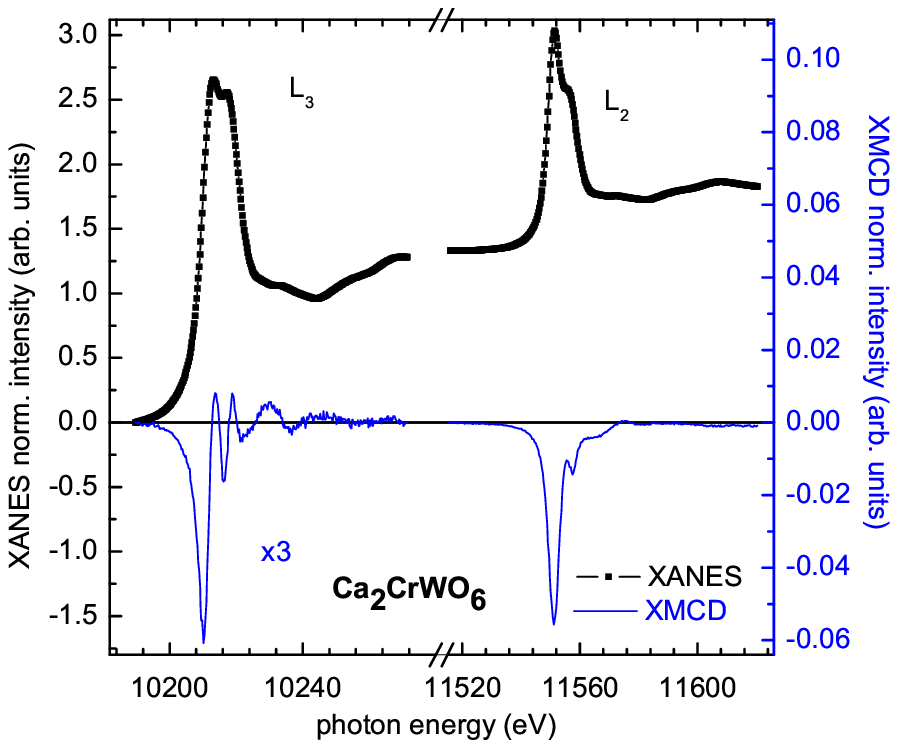}}
 \caption{(Color online) XANES spectra and derived XMCD spectra for Sr$_2$CrWO$_6$ (upper panel) and
 Ca$_2$CrWO$_6$ (lower panel). The XANES spectra (symbols) are corrected as
 described in the text. The XMCD spectra are shown as full lines.
 The XMCD spectrum of the W $L_3$ edge was multiplied by a factor
 7 for Sr$_2$CrWO$_6$ and by a factor of 3 for Ca$_2$CrWO$_6$ for clarity.}
 \label{Fig:XMCD}
\end{figure}

We first discuss the XANES spectra of Sr$_2$CrWO$_6$ and
Ca$_2$CrWO$_6$ shown in Fig.~\ref{Fig:XMCD}. As expected, a
similar behavior is found for the closely related compounds. The
white lines at the W $L_{2,3}$ edges have a rich fine structure
which is related to the valency and crystal field \cite{Besse:02}.
The $L_3$ absorption edges showed a clear double peak structure
with slightly less intensity of the peak at higher energy, whereas
at the $L_2$ edge this peak is less intense and forms a high
energy shoulder. This double peak structure is identified as the
signature of the crystal field splitting ($\sim3.2$\,eV for
Sr$_2$CrWO$_6$) of the $5d$ band into $t_\textrm{2g}$ and
$e_\textrm{g}$ states. Similar double peak structures have also
been observed at the Mo $L_{2,3}$ absorption edges for the double
perovskite Sr$_2$FeMoO$_6$ \cite{Besse:02}, however, with a more
pronounced separation between the peaks and slightly different
intensities at the $L_3$ edge.

\begin{table}[tb]
\caption{\label{tab:table1} Measured (exp., normalized to 5\,K)
and calculated (th.) magnetic moments per formula unit (f.u.) at
the 'non-magnetic' ions (W, Mo, Re) for different double
perovskites at 5\,K.}
\begin{ruledtabular}
\begin{tabular}{llccc}
 & material & $m_{\textrm{S}}$ ($\mu_{\textrm{B}}$/f.u.) & $m_{\textrm{L}}$ ($\mu_{\textrm{B}}$/f.u.) & $|m_{\textrm{L}}/m_{\textrm{S}}|$
 \\ \hline
exp. & Ca$_2$CrWO$_6$ & $-0.22\pm0.02$ & $0.10\pm0.01$ & $0.44\pm0.03$\\
 & Sr$_2$CrWO$_6$ & $-0.33\pm0.02$ & $0.12\pm0.02$ & $0.35\pm0.01$              \\
 & Sr$_2$FeMoO$_6$ \cite{Besse:02}& $-0.32\pm0.05$ & $-0.05\pm0.05$ & $0.15$     \\
 \hline
th. & Ca$_2$CrWO$_6$ & -0.25 & 0.06 & 0.25 \\
 & Sr$_2$CrWO$_6$ & -0.31 & 0.10 & 0.32\\
 & Sr$_2$FeMoO$_6$ \cite{Kanchana:05} & -0.24 & 0.02 & 0.09 \\
 & Sr$_2$CrReO$_6$ \cite{Vaitheeswaran:05} & -0.85 & 0.18 & 0.21\\
\end{tabular}
\end{ruledtabular}
\end{table}

As shown in Fig.~\ref{Fig:XMCD}, for both absorption edges we find
a rather intense XMCD signal. This is a clear evidence for the
existence of a magnetic moment at the W $5d$ shell. While the XMCD
spectra at the $L_2$ edge in Sr and Ca compounds appear to be very
similar, there are distinct differences between the $L_3$ XMCD
curves of Sr$_2$CrWO$_6$ and Ca$_2$CrWO$_6$. It is first very
interesting to note that the XMCD at the $L_3$ edge in
Ca$_2$CrWO$_6$ is nearly two times larger than in Sr$_2$CrWO$_6$.
Second, for Ca$_2$CrWO$_6$ the first minimum is significantly
stronger than for Sr$_2$CrWO$_6$, whereas for Sr$_2$CrWO$_6$ the
first maximum is more pronounced. These differences can be
attributed to different radii of the $A$ site ions, which cause
changes in the crystal and, in turn, the electronic and magnetic
structure. Third, the observed XMCD oscillations for both
compounds are strongly damped (but still present) at the $L_2$
edge as compared to the $L_3$ edge. This shows that the probed
orbital and spin resolved density of states is not identical
\cite{Ravindran:01}.

In Table~\ref{tab:table1} we list the magnetic moments at the W
site derived from the XMCD measurements using the standard sum
rules \cite{Thole:92,Carra:93} and compare them to theoretical
values. For completeness, we include also data previously
published by other groups on similar compounds
\cite{Besse:02,Vaitheeswaran:05,Kanchana:05}. Furthermore, the
ratio $\mid m_{\textrm{L}}/m_{\textrm{S}}\mid$ is calculated,
since this quantity is not affected by possible uncertainties in
the calculated number of holes. Table~\ref{tab:table1} shows that
the $5d$ magnetic spin moment of W
$m_{\textrm{S}}=-0.33\,\mu_{\textrm{B}}$/f.u. derived for
Sr$_2$CrWO$_6$ is in excellent agreement with the theoretically
calculated value. We note that this degree of agreement is a bit
fortuitous, since effects of anti-site disorder and the exact
oxygen content of the investigated samples have not been taken
into account in the theoretical analysis. We further note that the
orbital moment is only three times smaller than the spin moment.
From this we can conclude that spin-orbit coupling of the
delocalized $t_\textrm{2g}$ electron at the W site is of
considerable size, although not able to destroy the half metallic
character of this compound.


Comparing the magnetic spin moment at the W atom for the two
compounds, we find a reduction of the magnetic spin moment of only
$m_\textrm{S}=-0.22\,\mu_\textrm{B}$/f.u. and also a reduced
orbital moment for Ca$_2$CrWO$_6$ compared to Sr$_2$CrWO$_6$. This
corresponds to the structural differences between these compounds.
On replacing Sr$^{2+}$ by the smaller ion Ca$^{2+}$, the tolerance
factor $f$ \cite{Goodenough:rev} of the $A_2$CrWO$_6$ system
deviates significantly from unity and the crystal structure
changes from cubic to monoclinic. It has been pointed out recently
that for all ferrimagnetic double perovskites $T_\textrm{C}$ seems
to be maximum for $f\simeq 1$, i.e.~for the undistorted cubic
structure \cite{Philipp:03}. The only exception from this rule is
Ca$_2$FeReO$_6$ \cite{Westerburg:02}. Due to the monoclinic
distortion in Ca$_2$CrWO$_6$ the $B$-O-$B^\prime$ bonding angle
deviates significantly from $180^\circ$. This results in a
reduction of the hopping integral, i.e. in a weaker delocalization
of the W 5$d$ electron and, in turn, in a weakening of the
magnetic exchange. As a direct consequence, $T_\textrm{C}$ and the
magnetic moment induced at the W are reduced in Ca$_2$CrWO$_6$.
We note that a large magnetic
moment is theoretically predicted for the Re site in
Sr$_2$CrReO$_6$. However, so far no experimental verification is
made by XMCD measurements.


We finally discuss our findings in the context of the model
developed by Sarma {\em et al.}~\cite{Sarma:00} as described in
short above. The key point of this model is that the ferromagnetic
coupling between the magnetic ions is established by (fully) spin
polarized charge carriers with opposite magnetization direction
originating from the 'non-magnetic' site. The spin magnetic moment
is smeared out over several sites and bonds, however, it must lead
to a finite negative spin magnetization at the 'non-magnetic'
site. Previous experiments \cite{Besse:02} and our experiment here
show that in double perovskites with similar $T_\textrm{C}$, also
similar magnetic moments are found at the 'non-magnetic' site (Mo
in Sr$_2$FeMoO$_6$ and W in Sr$_2$CrWO$_6$, see
Table~\ref{tab:table1}). As soon as the kinetic exchange is
reduced due to lattice distortions, the spin magnetic moment
decreases as shown in the case of Ca$_2$CrWO$_6$, at the same time
also $T_\textrm{C}$ decreases. In this sense, the magnetic
coupling, the degree of delocalization, and the magnetic moment at
the 'non-magnetic' site are consequences of the kinetic energy
driven exchange model. However at this stage, it is difficult to
quantify these relations or establish a theory to calculate
$T_\textrm{C}$. It would be also important to confirm the
half-metallicity of the double perovskites in question by
additional methods as spin-resolved photoemission spectroscopy or
tunneling magneto-resistance effects.

In summary, we have performed XMCD measurements of the magnetic
moment at the $5d$ shell of the 'non-magnetic' ion W in the double
perovskite system $A_2$CrWO$_6$ with $A=$\,Sr, Ca. Our
experimental results are in good agreement with those of our
band-structure calculations. The smaller W $5d$ magnetic moment
found for the Ca$_2$CrWO$_6$ compound is attributed to an enhanced
localization of the W $5d$ electron due to a mono\-clinic
distortion. Our results suggest that there is a correlation
between the magnitude of the magnetic moment at the non-magnetic
ion and the magnetic ordering temperature $T_\textrm{C}$ in the
double perovskites. We believe that this correlation can provide
important clues on how to construct an accurate theory for the
magnetic ordering temperature in double perovskites.

This work was supported by the Deutsche Forschungsgemeinschaft, by
the BMBF (project 13N8279) and by the ESRF (HE-1658).

\end{document}